\documentclass[conference]{IEEEtran}
\IEEEoverridecommandlockouts
\usepackage{float}
\usepackage{cite}
\usepackage{amsthm,amsmath,amssymb,amsfonts}
\usepackage{algorithm,algorithmic}
\usepackage{graphicx}
\usepackage{textcomp}
\usepackage{xcolor}
\usepackage{tikz}
\usepackage{booktabs}

\usetikzlibrary{arrows,shapes,chains}
\usepackage[colorlinks,
            linkcolor=blue,
            anchorcolor=blue,
            citecolor=blue]{hyperref}
\hypersetup{
	colorlinks,
	citecolor=black,
	filecolor=black,
	linkcolor=black,
	urlcolor=black
}

\newtheorem{thm}{Theorem}

\newtheorem{lem}{Lemma}

\newtheorem{rem}{Remark}
\newtheorem{asum}{Assumption}

\newcommand{\tabincell}[2]{
	\begin{tabular}{@{}#1@{}}#2\end{tabular}
}

\def\BibTeX{{\rm B\kern-.05em{\sc i\kern-.025em b}\kern-.08em
    T\kern-.1667em\lower.7ex\hbox{E}\kern-.125emX}}
\begin{document}

\title{Lattice piecewise affine approximation of explicit nonlinear model predictive control with application to trajectory tracking of mobile robot*\\
\thanks{This work was supported in part by the National Natural Science Foundation of China under Grant U1813224, 62173113, and in part by the Science and Technology Innovation Committee of Shenzhen Municipality under Grant GXWD20201230155427003-20200821173613001, JCYJ20200109113412326, and in part by Natural Science Foundation of Guangdong Province of China under Grant 2022A1515011584.}
}

\author{\IEEEauthorblockN{1\textsuperscript{st} Kangbo Wang}
\IEEEauthorblockA{\textit{Harbin Institute of Technology} \\
Shenzhen, China \\
1960358110@qq.com}
\and
\IEEEauthorblockN{2\textsuperscript{nd} Kaijie Zhang}
\IEEEauthorblockA{\textit{Harbin Institute of Technology} \\
Shenzhen, China \\
harry6190@163.com}
\and
\IEEEauthorblockN{3\textsuperscript{nd} Yating Hang}
\IEEEauthorblockA{\textit{Harbin Institute of Technology} \\
Shenzhen, China \\
20s153112@stu.hit.edu.cn}
\and
\IEEEauthorblockN{4\textsuperscript{rd} Jun Xu}
\IEEEauthorblockA{\textit{Harbin Institute of Technology} \\
Shenzhen, China \\
xujunqgy@hit.edu.cn}
}

\maketitle

\begin{abstract}
To promote the widespread use of mobile robots in diverse fields, the performance of trajectory tracking must be ensured. To address the constraints and nonlinear features associated with mobile robot systems, we apply nonlinear model predictive control (MPC) to realize the trajectory tracking of mobile robots.
Specifically, to alleviate the online computational complexity of nonlinear MPC, this paper devises a lattice piecewise affine (PWA) approximation method that can approximate both the nonlinear system and control law of explicit nonlinear MPC. The kinematic model of the mobile robot is successively linearized along the trajectory to obtain a linear time-varying description of the system, which is then expressed using a lattice PWA model. Subsequently, the nonlinear MPC problem can be transformed into a series of linear MPC problems. Furthermore, to reduce the complexity of online calculation of multiple linear MPC problems, we approximate the optimal solution of the linear MPC by using the lattice PWA model. That is, for different sampling states, the optimal control inputs are obtained, and lattice PWA approximations are constructed for the state control pairs. Simulations are performed to evaluate the performance of our method in comparison with the linear MPC and explicit linear MPC frameworks. The results show that compared with the explicit linear MPC, our method has a higher online computing speed and can decrease the offline computing time without significantly increasing the tracking error. %The nonlinear system subjected to the lattice PWA approximated controller is proved to be stable in certain conditions. 
\end{abstract}

\begin{IEEEkeywords}
explicit nonlinear MPC, successive linearization, lattice piecewise affine approximation
\end{IEEEkeywords}

\section{Introduction}
Due to scientific advancements, mobile robots, especially wheeled mobile robots (WMRs), have been widely used in the military, exploration and other fields as well as dangerous operations. The task execution ability and intelligence of mobile robots depend on their trajectory tracking performances. With the development of control technology, mobile robots need to perform more difficult tasks in more demanding environments, which imposes stricter requirements on the accuracy and speed of trajectory tracking.

The pioneering research on the trajectory tracking of WMRs was performed by \cite{wang2016design}, who designed a Proportional-Integral-Derivative tracking controller to track the reference speed of the mobile robot. Subsequently, \cite{kchaou2019robust} combined the integral sliding mode surface with the adaptive observer to design an output feedback sliding mode controller. In addition, \cite{chu2017adaptive} designed an adaptive neural sliding-mode controller based on backstepping control and sliding mode control. However, existing algorithms do not consider the various constraints existing in actual systems. Thus, the model predictive control (MPC) algorithm has been widely applied for trajectory tracking owing to its ability to solve optimization problems with constraints.

The MPC algorithm predicts the future dynamic behavior of the system based on the model. By adding constraints on the future input, output or state variables, the constraints can be explicitly expressed in a programming problem solved online, as shown in \cite{maciejowski2002predictive,garcia1989model}. For example, \cite{conceiccao2007nonlinear} used the MPC method to design the trajectory tracking controller of a mobile robot with a nonlinear kinematic model. In addition, the continuous linearization method was proposed in \cite{kunhe2005mobile} to alleviate the large computational burden associated with nonlinear MPC. Moreover, explicit MPC has emerged as a promising strategy to reduce online computational complexity.

According to \cite{bemporad2002explicit,bemporad2002model}, the explicit state feedback solutions of quadratic optimal control problems for discrete linear time-invariant systems subjected to constraints can be obtained in advance. Therefore, online computation can be transformed into a simple table-lookup process, which is explicit MPC. This method can accelerate the online computation and expand the application scenarios of MPC.

Most technical processes are nonlinear in nature, and explicit solutions, which are characterized by high computational efficiency and verifiability, are appropriate for addressing nonlinear MPC problems. For example, \cite{seo2021nonaffine} applied explicit nonlinear MPC to construct a control design framework for the optimal trajectory tracking of small helicopters based on a multi-time scale structure. Explicit multi-parametric nonlinear MPC relies on multi-parametric nonlinear programming (mpNLP) algorithms to derive control laws. However, the solution of mpNLP problems is extremely complex, and it is challenging to identify an exact solution. To address these limitations, \cite{johansen2002multi} proposes a numerical algorithm to approximate mpNLP for nonlinear systems, and locally approximates mpNLP problems through multi-parametric quadratic program (mpQP) solutions in each partition. In \cite{maurovic2011explicit}, the nonlinear model of the mobile robot was approximated by a continuous linear time-varying model, and the explicit linear MPC problem was solved in each discrete time to obtain the trajectory tracking control law.

Although explicit MPC algorithms can facilitate online computation, the numbers of state partitions and control laws increase dramatically with increasing problem complexity, which increases the required storage space and online lookup time. To alleviate online computational complexity, the lattice piecewise affine (PWA) model can be used to represent the optimal offline-calculated control rules \cite{wen2009analytical}. \cite{xu2016irredundant} specified the necessary and sufficient conditions and related algorithms for irredundant lattice PWA models, which were applied to express the solutions of explicit linear MPC problems. Subsequently, \cite{martinez2014digital} implemented the lattice PWA model using very-large-scale integrated circuits, which enhanced the computing speed and decreased the resource consumption.

In this study, the lattice PWA model is used both in approximating the nonlinear dynamics of the mobile robot and optimal control laws of explicit linear MPCs. First, we perform successive linearization along the trajectory of the mobile robot and a lattice PWA approximation model is constructed, based on which the global approximation error is defined. Second, for each explicit linear MPC problem, we obtain a lattice PWA approximation for the obtained control laws to prevent the division of all of the critical regions in the traditional explicit MPC. This step accelerates offline and online calculations and alleviates the computational complexity while ensuring the tracking accuracy.

The remainder of this paper is structured as follows. Section \text {II} describes the modeling of the WMR and successive linearization of the model along the trajectory. Section \text {III} describes the process of solving explicit MPC problems offline and the lattice approximation of the optimal control laws. Additionally, the online evaluation method is introduced. Section \text {IV} presents the comparative simulation results of linear MPC, explicit linear MPC and our method to demonstrate the advantages of our method in trajectory tracking applications. Section \text {V}  ends the paper with the concluding remarks.

\section{Formulation of trajectory tracking problem of the WMR based on successive linearization}\label{section2}
The trajectory tracking problem of the WMR is formulated as a nonlinear MPC problem, and the nonlinear kinematic model of the WMR is successively linearized to obtain successive linear MPC problems.

\subsection{Kinematic model of the WMR}\label{subsection2.1}
Fig. \ref{Fig1} shows the kinematic model of the WMR. In the inertial coordinate system OXY, the relevant variables of the kinematic model are the axis coordinates of the rear and front axle, denoted as $\left(X_{r}, Y_{r}\right)$ and $\left(X_{f}, Y_{f}\right)$, respectively. The center speed of the rear and front axle are denoted as $v$ and $v_{f}$. The deflection angle of the front wheel, the heading angle of the body and the wheelbase of the vehicle are respectively denoted as $\delta_{f}$, $\varphi$ and $l$. For convenience, the front and rear axles are not distinguished by subscripts in the remainder of this paper. According to the geometric relationship shown in Fig. \ref{Fig1}, the kinematic model of the WMR can be defined as follows:
		\begin{figure}[h]
			\centering
			\includegraphics[width=0.35\textwidth]{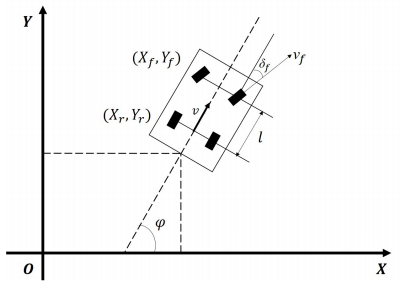} %1.png是图片文件的相对路径
			\caption{The kinematics model of WMR} %caption是图片的标题
			\label{Fig1} %此处的label相当于一个图片的专属标志，目的是方便上下文的引用
		\end{figure}

\begin{equation}
\begin{array}{ll}
\boldsymbol{\dot{\xi}}=\left[\begin{array}{c}
\dot{x} \\
\dot{y} \\
\dot{\varphi}
\end{array}\right]=\left[\begin{array}{c}
\cos \varphi \\
\sin \varphi \\
\frac{\tan \delta}{l}
\end{array}\right] v=f(\boldsymbol{\xi}, \boldsymbol{u})
\label{kinematic}
\end{array}
\end{equation}	
where $\boldsymbol{\xi}=\left[x , y , \varphi\right]^{T}$ and $\boldsymbol{u}=\left[v , \delta\right]^{T}$ denote the state and control vectors of the robot, respectively. The reference trajectories considered herein are reachable trajectories, i.e., each point on the trajectories satisfies the kinematic equation. Therefore, the function of the reference trajectory can be expressed as follows:
\begin{equation}
\begin{array}{ll}
\boldsymbol{\dot{\xi}}_{r}=f\left(\boldsymbol{\xi}_{r}, \boldsymbol{u}_{r}\right)
\end{array}
\end{equation}
For convenience, we uniformly use $\boldsymbol{x}$ to represent the state vector $\boldsymbol{\xi}$. 

\subsection{Model linearization}\label{subsection2.2}
\subsubsection{Successive linearization of the kinematic model}\label{subsubsection2.2.1}
\begin{figure}[h]
	\centering
	\includegraphics[width=0.5\textwidth]{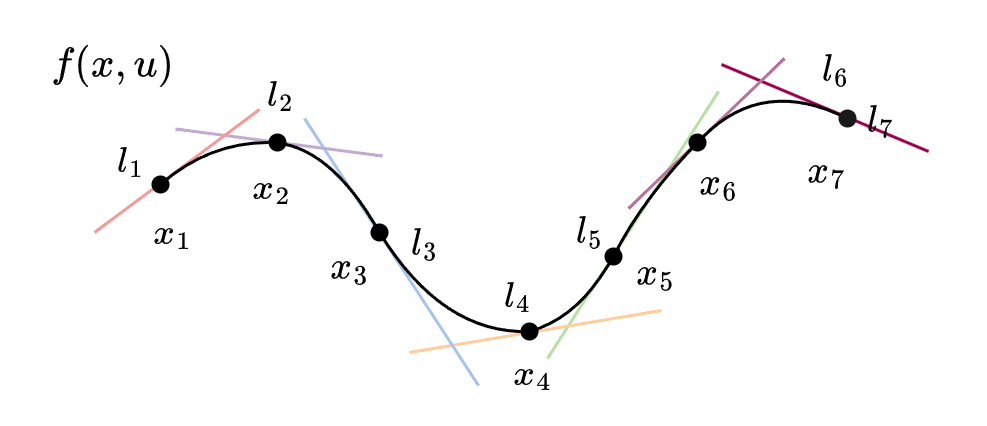} %1.png是图片文件的相对路径
	\caption{The sample trajectory points with different linearized models} %caption是图片的标题
	\label{linearize} %此处的label相当于一个图片的专属标志，目的是方便上下文的引用
\end{figure}

As the kinematic model of the WMR system is nonlinear, the resulting nonlinear MPC problem is difficult to solve. Thus, it's necessary to perform linearization. In a previous study, the continuous linearization method was applied to linearize a system model at the reference trajectory points corresponding to each sampling time \cite{maurovic2011explicit}. Similarly, in our study, the nonlinear kinematic model is linearized around all reference trajectory points, and the corresponding linear functions are obtained through Taylor expansion. These linear functions are connected to construct a lattice PWA model of the original nonlinear system, and then the approximation error is analyzed. Fig. $\ref{linearize}$ shows the lattice PWA approximation process for the function $f(\boldsymbol{x},\boldsymbol{u})$.

We sample the reference trajectory to be tracked at fixed time interval $T$ to obtain $K$ reference points $(\boldsymbol{x}_{k},\boldsymbol{u}_{k})$. The first-order Taylor expansion is applied to Equation (\ref{kinematic}) at each reference trajectory point, and the higher-order terms are ignored to obtain

\begin{equation}
\begin{array}{ll}
\dot{\boldsymbol{x}}\!\!\!\!&=\left.\frac{\partial f\!(\boldsymbol{x}, \boldsymbol{u})}{\partial \boldsymbol{x}}\right|_{\boldsymbol{x}\!=\!\boldsymbol{x}_{k}, \boldsymbol{u}\!=\!\boldsymbol{u}_{k}}\!\cdot\! \boldsymbol{x}\!+\!\left.\frac{\partial f\!(\boldsymbol{x}, \boldsymbol{u})}{\partial \boldsymbol{u}}\right|_{\boldsymbol{x}\!=\!\boldsymbol{x}_{k}, \boldsymbol{u}\!=\!\boldsymbol{u}_{k}}\!\cdot\! \boldsymbol{u}\!+\!f\!\!\left(\boldsymbol{x}_{k}, \boldsymbol{u}_{k}\right)\\
&-\left.\frac{\partial f(\boldsymbol{x}, \boldsymbol{u})}{\partial \boldsymbol{x}}\right|_{\boldsymbol{x}=\boldsymbol{x}_{k}, \boldsymbol{u}=\boldsymbol{u}_{k}}\cdot \boldsymbol{x}_{k}\!-\!\left.\frac{\partial f(\boldsymbol{x}, \boldsymbol{u})}{\partial \boldsymbol{u}}\right|_{\boldsymbol{x}=\boldsymbol{x}_{k}, \boldsymbol{u}=\boldsymbol{u}_{k}}\cdot \boldsymbol{u}_{k}\!
\label{taylor}
\end{array}
\end{equation}
Because the last three terms are related only to the state and control input of the reference trajectory, they can be regarded as constant terms. Therefore, after discretization through the first-order forward difference, Equation (\ref{taylor}) can be rewritten in the following concise form:
\begin{equation}
\begin{array}{ll}
\boldsymbol{x}(k+1)=A(k) \boldsymbol{x}(k)+B(k) \boldsymbol{u}(k)+b(k) 
\label{linearization}
\end{array}
\end{equation}

Assume that the number of reference points is $K$. Thus, after obtaining all of the discrete linear functions, we can construct a lattice PWA model (\ref{LTPWL}) to linearly approximate the original nonlinear system function in a piecewise fashion according to the algorithm proposed in \cite{wang2021lattice}.
\begin{equation}
\begin{array}{ll}
\hat{f}=f_{\text {LPWA}}=\max _{i=0, \ldots, K-1}\left\{\min _{j \in I_{\geq, i}}\left\{l_{j}\right\}\right\} \\
I_{\geq, i}=\left\{k \mid l_{k}\left(\boldsymbol{x}_{i}, \boldsymbol{u}_{i}\right) \geq l_{\operatorname{act}(i)}\left(\boldsymbol{x}_{i}, \boldsymbol{u}_{i}\right)\right\}
\label{LTPWL}
\end{array}                                                                    \end{equation}
where $l_{j}$ is the affine function defined in (\ref{linearization}), Specifically, we collect all the affine functions $[A(k) B(k) b(k)]\cdot [\boldsymbol{x}^T, \boldsymbol{U}^T, 1]^T$, identify the distinct ones and label them as $l_1, \ldots$. In this case, $l_{\operatorname{act}(i)}\left(\boldsymbol{x}_{i}, \boldsymbol{u}_{i}\right)$ 
is the activation function at point $(\boldsymbol x_i, \boldsymbol u_i)$. $I_{\geq, i}$ is the set of subscripts of affine functions. $\min _{j \in I_{\geq, i}}\left\{l_{j}\right\}$ is called term of the lattice PWA model and $s$ is the number of terms in the lattice model. The affine plane $l_{j}$ in each term is called the literal.

\subsubsection{Analysis of the global approximation error}\label{subsubsection2.2.2}
The following assumptions are made to analyze the error between the original and approximated system. For a nonlinear function $f(\boldsymbol{x},\boldsymbol{u})$, an approximate continuous PWA function $\hat{f}(\boldsymbol{x},\boldsymbol{u})$ can be constructed by linearizing the function at each reference point. The global approximation error can be obtained through the lattice PWA approximation of the kinematic model of the WMR.

\begin{asum}\label{Assumption1}
Given that $l_{\mathrm{act}(i)}(\boldsymbol{x},\boldsymbol{u})$ is the activation function of $f(\boldsymbol{x},\boldsymbol{u})$ at the linearization point $(\boldsymbol{x}_{i},\boldsymbol{u}_{i})$, then $f\left(\boldsymbol{x}_{i},\boldsymbol{u}_{i}\right)=l_{\mathrm{act}(i)}\left(\boldsymbol{x}_{i},\boldsymbol{u}_{i}\right)$, assume $\min _{j \in I_{\geq, i}}\left\{l_{j}\right\} \leq l_{i}\left(\boldsymbol{x}_{i},\boldsymbol{u}_{i}\right)$
\end{asum}
It has been proved in \cite{xu2021error} that if Assumption \ref{Assumption1} holds, we have
\begin{equation}
\begin{array}{ll}
\hat{f}\left(\boldsymbol{x}_{i},\boldsymbol{u}_{i}\right)=f\left(\boldsymbol{x}_{i},\boldsymbol{u}_{i}\right).
\end{array}                                                                                              
\end{equation}

\begin{asum}\label{Assumption2}
The original system $f(\boldsymbol{x},\boldsymbol{u})$ is Lipschitz continuous, i.e., $\forall (\boldsymbol x_1, \boldsymbol u_1), (\boldsymbol x_2, \boldsymbol u_2)$ in the domain of $f$, we have
\[
\|f(\boldsymbol x_1, \boldsymbol u_1)-f(\boldsymbol x_2, \boldsymbol u_2)\|\leq L_1\|x_1-x_2\|,
\]
in which $L_1$ is the Lipschitz constant.
\end{asum}

Under these two assumptions, the error between the original and approximated system follows Theorem \ref{Theorem1}.
\begin{thm}\label{Theorem1}
Suppose Assumption \ref{Assumption1} and \ref{Assumption2} hold. Assume that for any $(\boldsymbol{x},\boldsymbol{u}) \in \Omega$, in which $\Omega$ is the domain of $f(\boldsymbol{x},\boldsymbol{u})$, there is any sample point $(\boldsymbol{x}_{i},\boldsymbol{u}_{i})$ such that $\left\|(\boldsymbol{x},\boldsymbol{u})\!-\!(\boldsymbol{x}_{i},\boldsymbol{u}_{i})\right\| \!\leq\! \sigma$, we have $\|f(\boldsymbol{x},\boldsymbol{u})\!-\!\hat{f}(\boldsymbol{x},\boldsymbol{u})\|\! \leq\! L \sigma,\forall (\boldsymbol{x},\boldsymbol{u}) \in \Omega $, where $L$ is a constant related to $f$ and $\hat{f}$.
\end{thm}

\begin{proof} We first show that continuous PWA functions are Lipschitz continuous. For any two points $(\boldsymbol{x},\boldsymbol{u})$ and $(\boldsymbol{x}_{i},\boldsymbol{u}_{i})$, assume that the line segment $[(\boldsymbol{x},\boldsymbol{u}), (\boldsymbol{x}_{i},\boldsymbol{u}_{i})]$ intersect the linear subregion of $\hat{f}$ at points $(a_1, b_1), \ldots, (a_z, b_z)$, as shown in Fig. \ref{intersection}. Here linear subregions refer to regions where the continuous PWA function admits an affine expression.

\begin{figure}[h]
	\centering
	\includegraphics[width=0.5\textwidth]{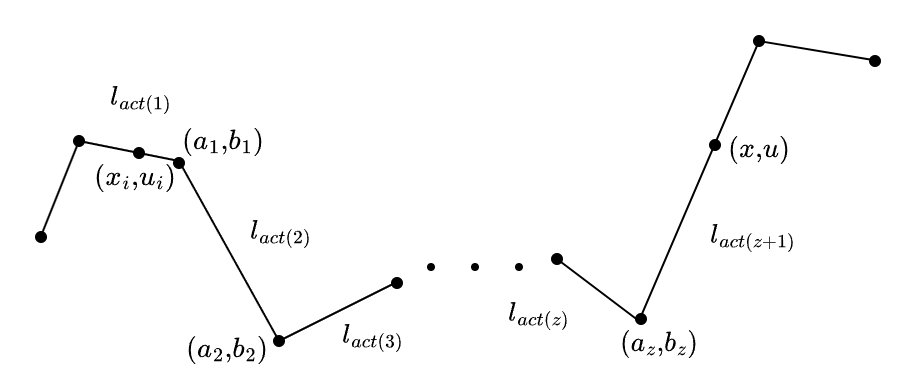} %1.png是图片文件的相对路径
	\caption{The intersection points and linear subregions} %caption是图片的标题
	\label{intersection} %此处的label相当于一个图片的专属标志，目的是方便上下文的引用
\end{figure}

In this case, the following expression holds.
\begin{equation}
\begin{array}{ll}
&||\hat{f}\!\left(\boldsymbol{x}_{i},\boldsymbol{u}_{i}\right)-\hat{f}(\boldsymbol{x},\boldsymbol{u})||\\ &\leq||\hat{f}\left(\boldsymbol{x}_{i},\boldsymbol{u}_{i}\right)-\hat{f}(a_{1},b_{1})+\hat{f}(a_{1},a_{1})\\
&\quad-\hat{f}(a_{2},b_{2})+\cdots+\hat{f}(a_{z}, b_{z})-\hat{f}(\boldsymbol{x},\boldsymbol{u})|| \\
&=|| l_{\mathrm{act}(1)}\left(\boldsymbol{x}_{i},\boldsymbol{u}_{i}\right)-l_{\mathrm{act}(1)}(a_{1},b_{1})+l_{\mathrm{act}(2)}(a_{1},b_{1})\\
&\quad-l_{\mathrm{act}(2)}(a_{2},b_{2})+\cdots+l_{\mathrm{act}(z)}(a_z, b_z)-l_{\mathrm{act}(z)}(\boldsymbol{x},\boldsymbol{u}) ||\\
&\leq M_{1}\left\|(\boldsymbol{x}_{i},\boldsymbol{u}_{i})-(a_{1},b_{1})\right\|+M_{2}\|(a_{1},b_{1})-(a_{2},b_{2})\|\\
&\quad+\cdots+M_{n}\|(a_{z}, b_{z})-(\boldsymbol{x},\boldsymbol{u})\| \\
&\leq N \cdot M_{\text {max }}\left\|\boldsymbol{x}_{i}-\boldsymbol{x}\right\|
\end{array}                                                                                              
\end{equation}
where $l_{\mathrm{act}(i)}(a_i,b_i)=l_{\mathrm{act}(i+1)}(a_i, b_i), i=1, \ldots, z-1$. Let $L_{2}=N \cdot M_{\max }$, so $\left\|\hat{f}\left(\boldsymbol{x}_{i},\boldsymbol{u}_{i}\right)-\hat{f}(\boldsymbol{x},\boldsymbol{u})\right\| \leq L_{2} \left\|\left(\boldsymbol{x}_{i},\boldsymbol{u}_{i}\right)-(\boldsymbol{x},\boldsymbol{u})\right\|$

When Assumption \ref{Assumption1} holds, according to Lipschitz continuity, we have
\begin{equation}
\begin{array}{ll}
&\left\|f(\boldsymbol{x},\boldsymbol{u})-\hat{f}(\boldsymbol{x},\boldsymbol{u})\right\|\\ &=\left\|f(\boldsymbol{x},\boldsymbol{u})-f\left(\boldsymbol{x}_{i},\boldsymbol{u}_{i}\right)+\hat{f}\left(\boldsymbol{x}_{i},\boldsymbol{u}_{i}\right)-\hat{f}(\boldsymbol{x},\boldsymbol{u})\right\| \\
& \leq L_{1}\left\|(\boldsymbol{x},\boldsymbol{u})-(\boldsymbol{x}_{i},\boldsymbol{u}_{i})\right\|+L_{2}\left\|(\boldsymbol{x},\boldsymbol{u})-(\boldsymbol{x}_{i},\boldsymbol{u}_{i})\right\| \\
&=\left(L_{1}+L_{2}\right)\left\|(\boldsymbol{x},\boldsymbol{u})-(\boldsymbol{x}_{i},\boldsymbol{u}_{i})\right\|
\end{array}                                                                                              
\end{equation}
If $\left\|(\boldsymbol{x},\boldsymbol{u})\!-\!(\boldsymbol{x}_{i},\boldsymbol{u}_{i})\right\| \!\leq\! \sigma$ and $L\!=\!L_{1}+L_{2}$, then $\|f(\boldsymbol{x},\boldsymbol{u})\!-\!\hat{f}(\boldsymbol{x},\boldsymbol{u})\|\! \leq\! L \sigma\!\!$. Therefore, provided that the distance between $(\boldsymbol{x},\boldsymbol{u})$ and $(\boldsymbol{x}_{i},\boldsymbol{u}_{i})$ is bounded by $\sigma$, the conclusion holds.
\end{proof}

We can change the error between the original and approximated system by adjusting the density of the linearization points, which can be achieved by changing the sampling time $T$.

\subsection{Trajectory tracking problem based on MPC}\label{subsection2.3}
The control objective of trajectory tracking is to ensure that the WMR rapidly attains the reference trajectory while satisfying the state, input and system model constraints, and keep the increment of control input as small as possible. Assuming the control and predictive horizons are both $N$, the MPC problem associated with trajectory tracking can be described as follows:
\begin{equation}
\begin{array}{ll}
\min _{U}\!\!\!&|J(k)=\sum_{k=0}^{N-1}\![(\boldsymbol{x}(k)-\boldsymbol{x}_{r}(k))^{\mathrm{T}} Q (\boldsymbol{x}(k)-\boldsymbol{x}_{r}(k))\!\\
&\qquad \quad \qquad \ \ +(\boldsymbol{u}(k)-\boldsymbol{u}_{r}(k))^{\mathrm{T}} R (\boldsymbol{u}(k)-\boldsymbol{u}_{r}(k))] |\\
s.t.&\boldsymbol{x}(k+1)=f(\boldsymbol{x}(k), \boldsymbol{u}(k)) \\
&\boldsymbol{u}_{\min }(k) \leq \boldsymbol{u}(k) \leq \boldsymbol{u}_{\max }(k), k=0, \ldots, N-1 \\
&\boldsymbol{x}_{\min }(k) \leq \boldsymbol{x}(k) \leq \boldsymbol{x}_{\max }(k), k=0, \ldots, N-1
\label{NMPC}
\end{array}
\end{equation}
where $f(\boldsymbol{x}(k), \boldsymbol{u}(k))$ is the nonlinear kinematic model described in (\ref{kinematic}); $\boldsymbol{u}_{\min }, \boldsymbol{u}_{\max }, \boldsymbol{x}_{\min }$ and $\boldsymbol{x}_{\max }$ are the lower and upper bounds of $\boldsymbol{u}$ and $\boldsymbol{x}$, respectively; and $Q=Q^{T} \succeq 0$ and $R=R^{T} \succ 0$ are the weighting matrices of the state and control terms, respectively.

In the subsection \ref{subsection2.2}, we have obtained the linear time-varying description of the kinematic model. Thus, the problem \ref{NMPC}) can be transformed to the following linear MPC problem at each sample trajectory point:
\begin{subequations}
\begin{align}
\min _{U}\!\!\!&|J(k)=\sum_{k=0}^{N-1}\![(\boldsymbol{x}(k)-\boldsymbol{x}_{r}(k))^{\mathrm{T}} Q (\boldsymbol{x}(k)-\boldsymbol{x}_{r}(k))\!\\
&\qquad \quad \qquad \ \ +(\boldsymbol{u}(k)-\boldsymbol{u}_{r}(k))^{\mathrm{T}} R (\boldsymbol{u}(k)-\boldsymbol{u}_{r}(k))] | \label{lmpc:cost}\\
s.t.& \boldsymbol{x}(k+1)=A(k) \boldsymbol{x}(k)+B(k) \boldsymbol{u}(k)+b(k) \label{lmpc:affinecon}\\
&\boldsymbol{u}_{\min }(k) \leq \boldsymbol{u}(k) \leq \boldsymbol{u}_{\max }(k), k=0, \ldots, N-1 \\
&\boldsymbol{x}_{\min }(k) \leq \boldsymbol{x}(k) \leq \boldsymbol{x}_{\max }(k), k=0, \ldots, N-1
\label{LMPC}
\end{align}
\end{subequations}
where $A(k)$, $B(k)$, and $b(k)$ are derived from Equation (\ref{linearization}).

\section{Lattice PWA approximation of the optimal controller of multiple linear MPC problems}\label{section3}
In subsection \ref{subsection2.3}, we describe the trajectory tracking problem of the WMR as a series of linear MPC problems. We find that if the prediction horizon is large or the state dimension is high, the online computation complexity of the resulting linear MPC problem is high. Therefore, the lattice PWA function is used to solve this problem, i.e., the optimal controller for linear MPC is calculated for several sample points and then approximated using the lattice PWA function. In this manner, the online calculation complexity can be further reduced.

\subsection{Main framework}\label{subsection3.1}
\begin{figure}[h]
	\centering
	\includegraphics[width=0.5\textwidth]{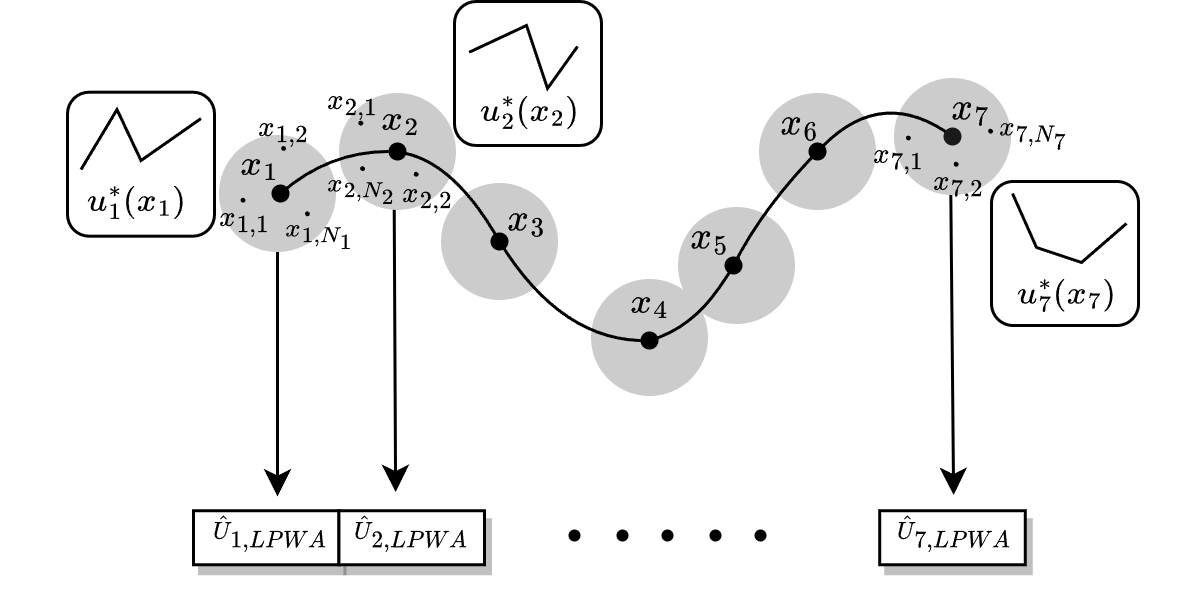} %1.png是图片文件的相对路径
	\caption{Linearization points and the corresponding lattice PWA approximation expressions} %caption是图片的标题
	\label{Table} %此处的label相当于一个图片的专属标志，目的是方便上下文的引用
\end{figure}

Consider the trajectory shown in Fig. \ref{Table} as an example. Assume that all of the corresponding linearized models are obtained as described in \ref{subsection2.2}. Firstly, for the first reference trajectory point $\boldsymbol{x}_{1}$, the corresponding optimization problem is shown in Equation (\ref{LMPC}). It has been shown in \cite{bemporad2002explicit} that optimal solution of the optimization problem (\ref{LMPC}), say $\boldsymbol u_1^*(\boldsymbol x_1)$ is a continuous PWA function of the state $\boldsymbol{x}_1$. As the process of obtaining such continuous PWA function is also time-consuming, we construct a lattice PWA approximation $\hat{\boldsymbol{u}}_{1,L\!P\!W\!A}$ for $u_1^*(\boldsymbol x_1)$. Specifically, sample points are generated around the point $\boldsymbol x_1$, e.g., $\boldsymbol x_{1,1}, \ldots, \boldsymbol x_{1, N_1}$, and the corresponding optimal solutions $u_1^*(\boldsymbol x_{1,1}), \ldots, u_1^*(\boldsymbol x_{1,N})$ are obtained. Next, the lattice PWA approximation $\hat{\boldsymbol{u}}_{1,L\!P\!W\!A}$ is formulated based on $(\boldsymbol x_{1,i}, \boldsymbol u_{1,i})_{i=1}^{N_1}$.

Subsequently, we formulate the linear MPC problem for the second reference trajectory point $\boldsymbol{x}_{2}$ by using (\ref{LMPC}). In this case, the cost function (\ref{lmpc:cost}) and constraint (\ref{lmpc:affinecon}) are different from those for the trajectory point $\boldsymbol x_1$, i.e., the reference point and corresponding linearized system are different. Sample point pairs $(\boldsymbol x_{2,i}, \boldsymbol u_{2,i})_{i=1}^{N_2}$ are generated for this newly constructed linear MPC problem in a similar manner, and the lattice PWA approximation  $\hat{\boldsymbol{u}}_{2,L\!P\!W\!A}$ is obtained to approximate $u_2^*(\boldsymbol x_2)$.

%The calculation steps for it are the same as those for $\boldsymbol{x}_{1}$, except that the parameters $H, F, C_{f}, G, W$ and $E$ in the optimization problem are calculated based on $\boldsymbol{x}_{2}$ and the linearized model $l_{2}$, and the sampling region is centered on $\boldsymbol{x}_{2}$. 

The abovementioned steps are repeated for the remaining trajectory points $\boldsymbol x_3, \ldots, \boldsymbol{x}_{7}$ to yield seven lattice PWA approximations $\hat{\boldsymbol{u}}_{1,L\!P\!W\!A}, \ldots, \hat{\boldsymbol{u}}_{7,L\!P\!W\!A}$.

%Finally, we save the information of the corresponding irredundant lattice PWA model $\hat{\boldsymbol{u}}_{2,L\!P\!W\!A}$. . So far, we have obtained a total of eight irredundant lattice PWA models.

When tracking online, we substitute the initial state point $\boldsymbol{x}_{real,1}$ of the WMR into the first lattice PWA expression to determine the first approximated control law $\boldsymbol{u}_{real,1}$:
\[
\boldsymbol{u}_{real,1}=\hat{\boldsymbol{u}}_{1,L\!P\!W\!A}(\boldsymbol x_{real, 1}).
\]
%$$
%		\boldsymbol{u}_{real1}\left(\boldsymbol{x}_{real1}\right)=\max _{i=1, \ldots, N_{1}}\left\{\min _{j=1, \ldots, N_{i}} \boldsymbol{u}_{j}\left(\boldsymbol{x}_{real1}\right)\right\}
%$$
The next state $\boldsymbol{x}_{real, 2}$ is obtained using $\boldsymbol{x}_{real,1}$ and $\boldsymbol{u}_{real,1}$ as
\[	\boldsymbol{x}_{real, 2}=f\left(\boldsymbol{x}_{real, 1}, \boldsymbol{u}_{real, 1}\right)
\]

Subsequently, the approximated control law $\boldsymbol{u}_{real, 2}$ and state $\boldsymbol x_{real, 3}$ are obtained in a similar manner. The abovementioned steps are repeated until the last time instant is reached. At this point, the actual trajectory of the mobile robot is fully defined.

To obtain the lattice PWA approximation $\hat{\boldsymbol{u}}_{1,L\!P\!W\!A}, \ldots,$, we determine $u^*(\boldsymbol x)$ with respect to $\boldsymbol x$ for a specific linear MPC problem and construct the lattice PWA approximation for the optimal solution of a linear MPC problem, as described in the following subsections.
% Additionally, the system stability under lattice PWA approximations is analyzed.

\subsection{Offline determination of the optimal solution of linear MPC}\label{subsection3.2}
To calculate the optimal solution of the linear MPC problem for the sample points, we adopt the idea of explicit MPC and express the optimal solution $\boldsymbol u^*(\boldsymbol x)$ as an affine function of $\boldsymbol x$.

%When the prediction horizon $N$ is large or the constraints are complex, it is time-consuming to solve (\ref{LMPC}) online, hence we resort to explicit MPC. 
$X(k) = [\boldsymbol{x}(k), ..., \boldsymbol{x}(k\!+\!N\!-\!1)]^{\mathrm{T}}$ and $U(k) = [\boldsymbol{u}(k), ..., \boldsymbol{u}(k\!+\!N\!-\!1)]^{\mathrm{T}}$ represent the sequence of state and control quantities from the current time $k$ to the future time $k+N-1$, respectively. $\boldsymbol x$ represents $\boldsymbol{x}(k)$, for simplicity. Now, the optimization problem (\ref{LMPC}) can be restated as the following mpQP problem:
\begin{equation}
\begin{array}{ll}
\min _{U} \frac{1}{2} U^{T} H U+\left(\boldsymbol{x}^{T} F+C_{f}\right) U \\
\text { s.t.}\quad G U \leq W+E \boldsymbol{x}
\label{mpqp}
\end{array}
\end{equation}
in which matrices $H,\ F,\ C_{f},\ G,\ W,\ E$ can be obtained through (\ref{LMPC}) . By defining $\boldsymbol{z}=U+H^{-1} F^{T} \boldsymbol{x}$, we can deform problem (\ref{mpqp}) into the following standard quadratic programming form (\ref{stand}).
\begin{equation}
\begin{array}{ll}
\begin{aligned}
V_{\boldsymbol{z}}(\boldsymbol{x})=& \min _{\boldsymbol{z}} \frac{1}{2} \boldsymbol{z}^{T} H \boldsymbol{z} \\
& \text { s.t. }  G \boldsymbol{z} \leq \omega+S \boldsymbol{x}
\label{stand}
\end{aligned}
\end{array}
\end{equation}

The Karush—Kuhn--Tucker (KKT) conditions for the standard quadratic programming problem are
\begin{subequations}
\begin{align}
H \boldsymbol{z}^{*}+G_{\mathcal{A}^{*}}^{T} \lambda^{*}+G_{\mathcal{N}^{*}}^{T} \mu^{*}=0 \label{Wa}\\
G_{\mathcal{A}^{*}} \boldsymbol{z}^{*}=\omega_{\mathcal{A}^{*}}+S_{\mathcal{A}^{*}} \boldsymbol{x} \label{Wb}\\
G_{\mathcal{N}^{*}} \boldsymbol{z}^{*}<\omega_{\mathcal{N}^{*}}+S_{\mathcal{N}^{*}} \boldsymbol{x} \label{Wc}\\
\lambda^{*} \geq 0 \label{Wd}\\
\mu^{*} \geq 0 \label{We}\\
\lambda^{* T}\left(G_{\mathcal{A}^{*}} \boldsymbol{z}^{*}-\omega_{\mathcal{A}^{*}}-S_{\mathcal{A}^{*}} \boldsymbol{x}\right)=0 \label{Wf}\\
\mu^{* T}\left(G_{\mathcal{N}^{*}} \boldsymbol{z}^{*}-\omega_{\mathcal{N}^{*}}-S_{\mathcal{N}^{*}} \boldsymbol{x}\right)=0 \label{Wg}
\end{align}
\end{subequations}
where (\ref{Wb}) and (\ref{Wc}) are the active and inactive constraints of the optimal solution $\boldsymbol{z}^{*}$, respectively. Assuming that $G_{j}, \omega_{j}, S_{j}$ represents line $j$ of $G, \omega, S$, the sets of indices for the active and inactive constraints can be represented as follows:
$$
\mathcal{A}^{*}=\left\{j \in\{1, \ldots, p\} \mid G_{j} \boldsymbol{z}^{*}=\omega_{j}+S_{j} \boldsymbol{x}\right\} 
$$
$$
\mathcal{N}^{*}=\left\{j \in\{1, \ldots, p\} \mid G_{j} \boldsymbol{z}^{*}<\omega_{j}+S_{j} \boldsymbol{x}\right\}
$$

For a fixed set of active constraints, if $G_{\mathcal{A}^{*}}$ is full row rank, we can obtain the optimal solution and critical region $CR_{i}$ corresponding to the active constraint set. Given that $\boldsymbol{z}=U+H^{-1} F^{T} \boldsymbol{x}$, we can obtain the explicit expression of the control sequence $U$ in this region with respect to the state quantity $\boldsymbol{x}$.
\begin{equation}
\begin{array}{ll}
{U_{i}}^{*}\left(\boldsymbol{x}\right)\!\!\!&=\!H^{-1} G_{\mathcal{A}^{*}}^{T}\left(G_{\mathcal{A}^{*}} H^{-1} G_{\mathcal{A}^{*}}^{T}\right)^{-1}\left(\omega_{\mathcal{A}^{*}}+S_{\mathcal{A}^{*}} \boldsymbol{x}\right)\\
&- H^{-1}\left(\boldsymbol{x}^{T} F+C_{f}\right)^{T}
\label{solution}
\end{array}
\end{equation}

The obtained $U_{i}^{*}\left(\boldsymbol{x}\right)$ is the optimal control sequence. Only the first term $\boldsymbol{u}_{i}^{*}\left(\boldsymbol{x}\right)$, which is also an affine function of $\boldsymbol x$, is considered. It is defined as follows:
\begin{equation}\label{eq:u*}
\boldsymbol u^*_i(x)=[\boldsymbol 1_{n_u}, \boldsymbol 0_{(N-1)\times n_u}]\cdot U_i^*(\boldsymbol x),
\end{equation}
where $\boldsymbol 1_{n_u}$ and $\boldsymbol 0_{(N-1)\times n_u}$ are the all-one and all-zero vectors, respectively, with lengths $n_u$ and $(N-1)\times n_u$.
% (\emph{have to explain the dimensionality of $u$})

\begin{rem}\label{Remark2}
If $ G_{\mathcal{A}^{*}} $ is not full row rank, the linear independence constraints qualification is violated. Assuming that the rank of $ G_{\mathcal{A}^{*}} $ is $r$, we can arbitrarily select $r$ independent constraints as the new active constraint set to simplify the constraints.
\end{rem}

\subsection{Lattice PWA approximation of the optimal solution of the linear MPC problem}\label{subsection3.3}
For the linear MPC problem corresponding to each linearized model, the dataset consisting of the state and control law information is generated by sampling and resampling procedure.

\subsubsection{Data sampling}\label{subsubsection3.3.1}

\begin{figure}[h]
	\centering
	\includegraphics[width=0.5\textwidth]{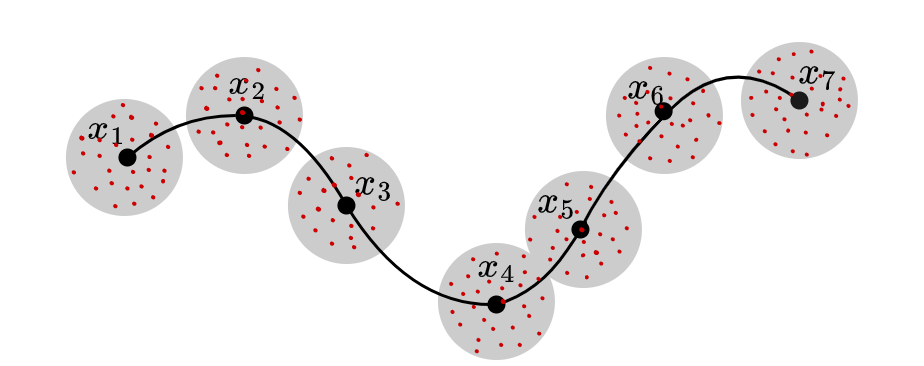} %1.png是图片文件的相对路径
	\caption{Sampling of state points around the linearization points.} %caption是图片的标题
	\label{point} %此处的label相当于一个图片的专属标志，目的是方便上下文的引用
\end{figure}

In \cite{xu2021error}, sample states are generated in the domain of interest. In the current study, because the actual state is near the reference points, the sample states are generated near the reference points $\boldsymbol x_1, \ldots$. Consider $\boldsymbol x_1$ as an example. We define a spherical region with radius $r$ centered on $\boldsymbol{x}_{1}$ and obtain the sample dataset $\mathcal{X}_{1} \times \mathcal{U}_{1}$ by sampling and resampling within the region. The value of $r$ can be determined by considering the distance between the actual and reference states. Provided that the sphere $\mathcal{B}(\boldsymbol x_1, r)$ covers all of the possible states, the lattice PWA approximations constructed offline can yield a solution close to the optimal solution of the linear MPC problem corresponding to $\boldsymbol x_1$. We first calculate the distance of all of the adjacent reference points, with the minimum distance being $d$. $r$ is determined to be $r=d/2$. Fig. \ref{point} illustrates the sampling of the state points around linearization points $\boldsymbol x_1, \ldots, \boldsymbol x_7$.

%Based on the dataset, the corresponding lattice PWA expression , and its terms and literals information are saved after simplification.
%. Assume the linearized model at time $k$ is $l_{k}$, we define a spherical region with radius  with the reference point $\boldsymbol{x}_{k}$ as the center of the sphere, and sample $N_{1}$ points $\boldsymbol{x}_{i}$ in the region, as shown in Fig. \ref{point} (In the figure, we only show a small number of sample points, when in fact we choose the number of sample points based on our requirement.). 

After sampling and resampling in $\mathcal{B}(\boldsymbol x_i, r)$, we obtain sample points $\boldsymbol x_{i,1}, \ldots, \boldsymbol x_{i, N_i}$ for each linearization point $\boldsymbol{x}_{i}$ on the reference trajectory. Next, we calculate the corresponding affine function according to (\ref{solution}). The obtained sample point $\left(\boldsymbol{x}_{i, j},\boldsymbol{u}_{i}^*\left(\boldsymbol{x}_{i, j}\right)\right)$ is stored in the $i\!$-th dataset $\mathcal{X}_{i} \times \mathcal{U}_{i}$.

\subsubsection{Lattice approximation}\label{subsubsection3.3.2}
After obtaining the sample dataset, the expression of lattice PWA can be derived based on $\mathcal{X}_{i} \times \mathcal{U}_{i}$:
\begin{equation}
\begin{array}{ll}
\hat{\boldsymbol{u}}_{i,L\!P\!W\!A}(\boldsymbol{x})=\max _{k=1, \ldots, N_{i}}\left\{\min _{j \in J_{\geq, i, k}} \boldsymbol{u}_{i, j}(\boldsymbol{x})\right\}
\label{lattice}
\end{array}
\end{equation}
where the index set $J_{\geq, i, k}$ is defined as
\[
J_{\geq, i, k}=\{j|\boldsymbol u_{i, j}(\boldsymbol x_{i, k})\geq \boldsymbol u_{i, k}(\boldsymbol x_{i, k})\},
\]
and the affine plane $\boldsymbol{u}_{j}(\boldsymbol{x})$ and $\min _{j \in J_{\geq, i}} \boldsymbol{u}_{j}(\boldsymbol{x})$ represent a literal and a term of the lattice PWA approximation, respectively. The affine plane $\boldsymbol u_{i, j}$ is obtained through (\ref{solution}) and (\ref{eq:u*}), and superscript $*$ is omitted for simplicity.

\cite{xu2021error} demonstrates that if all of the distinct affine functions are sampled, the approximation coincides with the optimal solution of the linear MPC problem.

\begin{lem}
Assume that all of the distinct affine functions are sampled. The following expression holds if the lattice PWA approximation is constructed using Equation (\ref{lattice}):
\label{lemma1}
\begin{equation}
\begin{array}{ll}
\hat{\boldsymbol{u}}_{i,L\!P\!W\!A}(\boldsymbol{x})=\boldsymbol{u}^{*}(\boldsymbol{x}), \forall \boldsymbol{x} \in \Gamma\left(\boldsymbol{x}_{i, k}\right), \forall \boldsymbol{x}_{i, k} \in \mathcal{X}_{i}
\end{array}
\end{equation}
where $\Gamma(\boldsymbol x_i)$ is the unique order region containing $\boldsymbol{x}_{i, k}$, i.e., the order of affine functions $u_{i, k}$ remains unchanged in $\Gamma(\boldsymbol x_i)$.
\end{lem}

In general, it is challenging to ensure the validity of the assumption in Lemma \ref{lemma1}. Therefore, \cite{xu2021error} proposed a resampling method to sample as more functions as possible. It is noted that as the sampling region around each linearization point is relatively small, it is not hard to guarantee that all the distinct affine functions in the sample region have been sampled. %should show in simulation

\subsubsection{Simplification}\label{subsubsection3.3.3}
The lattice PWA model obtained using this method will generate redundant parameters. To reduce the memory occupied by terms and literals and the complexity of online evaluation, we use the method proposed in \cite{xu2016irredundant} to remove both the redundant terms and literals of the lattice PWA model and save the remaining values.

The whole process for obtaining lattice PWA approximations is shown in Algorithm \ref{alg:1}.

\begin{algorithm}
	\renewcommand{\algorithmicrequire}{\textbf{Input:}}
	\renewcommand{\algorithmicensure}{\textbf{Output:}}
	\caption{Offline calculation}
	\label{alg:1}
	\begin{algorithmic}[1]
		\REQUIRE  Nonlinear kinematic model of WMR and $K$ reference trajectory points
		\ENSURE ${f}\!_{L\!P\!W\!A}(\boldsymbol{x}, \boldsymbol{u})$ and $\hat{\boldsymbol{u}}_{i, L\!P\!W\!A}(\boldsymbol{x}), i=1, \ldots, K$, as defined in (\ref{LTPWL}) and (\ref{lattice}) , respectively.
		\STATE Initialization: Prediction and control horizons, $N$; Number of state sample points, $N_{k}, k=1,\ldots, K$. 
		\FOR   {$m=1:K$}
		\STATE Obtain the linearized model $l_{m}(\boldsymbol{x}, \boldsymbol{u})$ by Taylor expansion at each sample point
		\ENDFOR
        \STATE Construct the lattice approximation model ${f}_{LPWA}(\boldsymbol{x}, \boldsymbol{u})$ $=\max _{m=1, \ldots, K}\left\{\min _{j \in I_{\geq, m}}\left\{l_{j}\right\}\right\}$
		\FOR   {$i=1:K$}
		\FOR   {$k=1:N_{k}$}
		\STATE Sample a point $x_{i, k}$ near the $k\!$-th reference trajectory point $\boldsymbol x_i$ and solve the mpQP problem to obtain $\boldsymbol{u}_{i, k}\left(\boldsymbol{x}_{i}\right)$
        \STATE Add sample point $\left(\boldsymbol{x}_{i, k}, \boldsymbol{u}_{i, k}\left(\boldsymbol{x}_{i, k}\right)\right)$ to dataset $\mathcal{X}_{i} \times \mathcal{U}_{i}$
		\ENDFOR
		\STATE Construct the lattice approximation model $\hat{\boldsymbol{u}}_{k,L\!P\!W\!A}(\boldsymbol{x})$ $=\max _{i=1, \ldots, N_{k}}\left\{\min _{j \in J_{\geq, i, k}} \boldsymbol{u}_{i, k}(\boldsymbol{x})\right\}$ based on $\mathcal{X}_{i} \times \mathcal{U}_{i}$.
      \STATE Simplification.
		\ENDFOR 
	\end{algorithmic}  
\end{algorithm}

\subsubsection{Online evaluation}\label{subsubsection3.3.4}
After obtaining the lattice PWA expressions corresponding to all reference points on the reference trajectory, we calculate the control laws online. The overall process of online evaluation is shown in Algorithm \ref{alg:2}.

\begin{algorithm}
	\renewcommand{\algorithmicrequire}{\textbf{Input:}}
	\renewcommand{\algorithmicensure}{\textbf{Output:}}
	\caption{Online evaluation}
	\label{alg:2}
	\begin{algorithmic}[1]
		\REQUIRE Lattice PWA approximations $\qquad$ $\hat{\boldsymbol{u}}_{i, L\!P\!W\!A}(\boldsymbol{x})$ corresponding to the $i$-th linear MPC problem and initial state $\boldsymbol x_{(1)}$ of the WMR
		\ENSURE Actual tracking state $x_{(k)}$ of the WMR
		\STATE Initialization: Number of reference track points: $K$ 
		\FOR   {$k=1:K$}
		\STATE Obtain the control law as $\hat{\boldsymbol{u}}_{k}(\boldsymbol{x}_{(k)})=\hat{\boldsymbol{u}}_{i, L\!P\!W\!A}(\boldsymbol{x}(k))$.
		\STATE Substitute  $\hat{\boldsymbol{u}}_{k}(\boldsymbol{x}_{k})$ into the origin kinematic model to obtain the next real state $\boldsymbol x_{(k+1)}$.
		\ENDFOR
	\end{algorithmic}  
\end{algorithm}

The control law evaluated online for the actual state value $\boldsymbol{x}_{(k)}$ at the $k\!$-th moment of the robot can be described as $\hat{\boldsymbol{u}}_{k}(\boldsymbol{x}_{(k)})=\hat{\boldsymbol{u}}_{i, L\!P\!W\!A}(\boldsymbol{x}(k))$.
$\boldsymbol{u}_{k}\left(\boldsymbol{x}_{k}\right)$ can be obtained by substituting $\boldsymbol{x}_{(k)}$ into the corresponding $k\!-\!th$ expression.

%For the actual state value $\boldsymbol{x}_{(k)}$ at the $k\!$-th moment of the robot, the control law online evaluated can be described as $\hat{\boldsymbol{u}}_{k}(\boldsymbol{x}_{(k)})=\hat{\boldsymbol{u}}_{i, L\!P\!W\!A}(\boldsymbol{x}(k))$.

Subsequently, the actual state value $\boldsymbol{x}_{(k+1)}$ that the robot reaches when driven by this control input can be determined using the system kinematic equation (\ref{kinematic}). In this manner, the WMR trajectory can be tracked online.

\section{Experimental Results}\label{section4}
Equation (\ref{kinematic}) is used as the kinematic model of the WMR with the following parameters: vehicle wheelbase $l = 0.1 m$, prediction and control horizons $N = 10$, and discrete sampling period $T = 0.1 s$. In general, trajectory tracking requires high accuracies for the horizontal and vertical coordinates of the mobile robot and low accuracy for the heading angle of the mobile robot. Therefore, the weight matrix of the objective function (\ref{LMPC}) is selected as follows:
\begin{equation}
\begin{array}{ll}
Q=\left[\begin{array}{ccc}
10 & 0 & 0 \\
0 & 10 & 0 \\
0 & 0 & 0.5
\end{array}\right], R=\left[\begin{array}{cc}
0.1 & 0 \\
0 & 0.1
\end{array}\right]
\end{array}
\end{equation}

To demonstrate the superiority of our method, identical parameter conditions are set for the nonlinear MPC, linear MPC and traditional explicit linear MPC.

\subsection{Circular reference trajectory}
We consider a circular reference trajectory and perform a trajectory tracking simulation experiment using the WMR. The initial state of the robot is
$\left[\begin{array}{llll}
	\!1.9\!\!& 0\!\!& 1.57\!
\end{array}\right]^{T}$.
Considering the actual situation of the WMR, the state and control quantity constraints are set as follows:
\begin{equation}
	\begin{array}{ll}
		\left[\begin{array}{c}
			-3.0 \\
			-3.0 \\
			-3 \pi
		\end{array}\right] \leq \boldsymbol{x} \leq\left[\begin{array}{c}
			3.0 \\
			3.0 \\
			3 \pi
		\end{array}\right], \left[\begin{array}{c}
			-2 \\
			-\frac{\pi}{2}
		\end{array}\right] \leq \boldsymbol{u} \leq\left[\begin{array}{c}
			2 \\
			\frac{\pi}{2}
		\end{array}\right]
	\end{array}
\end{equation}

Next, we linearize the kinematic model at each reference point and obtain 360 linearized models. Finally, for each explicit linear MPC problem, we sample 300 points near the reference point. The tracking performance of the three methods is evaluated in terms of the online computing time, offline computing time and average tracking error.

		\begin{figure}[h]
			\centering
			\includegraphics[width=0.5\textwidth]{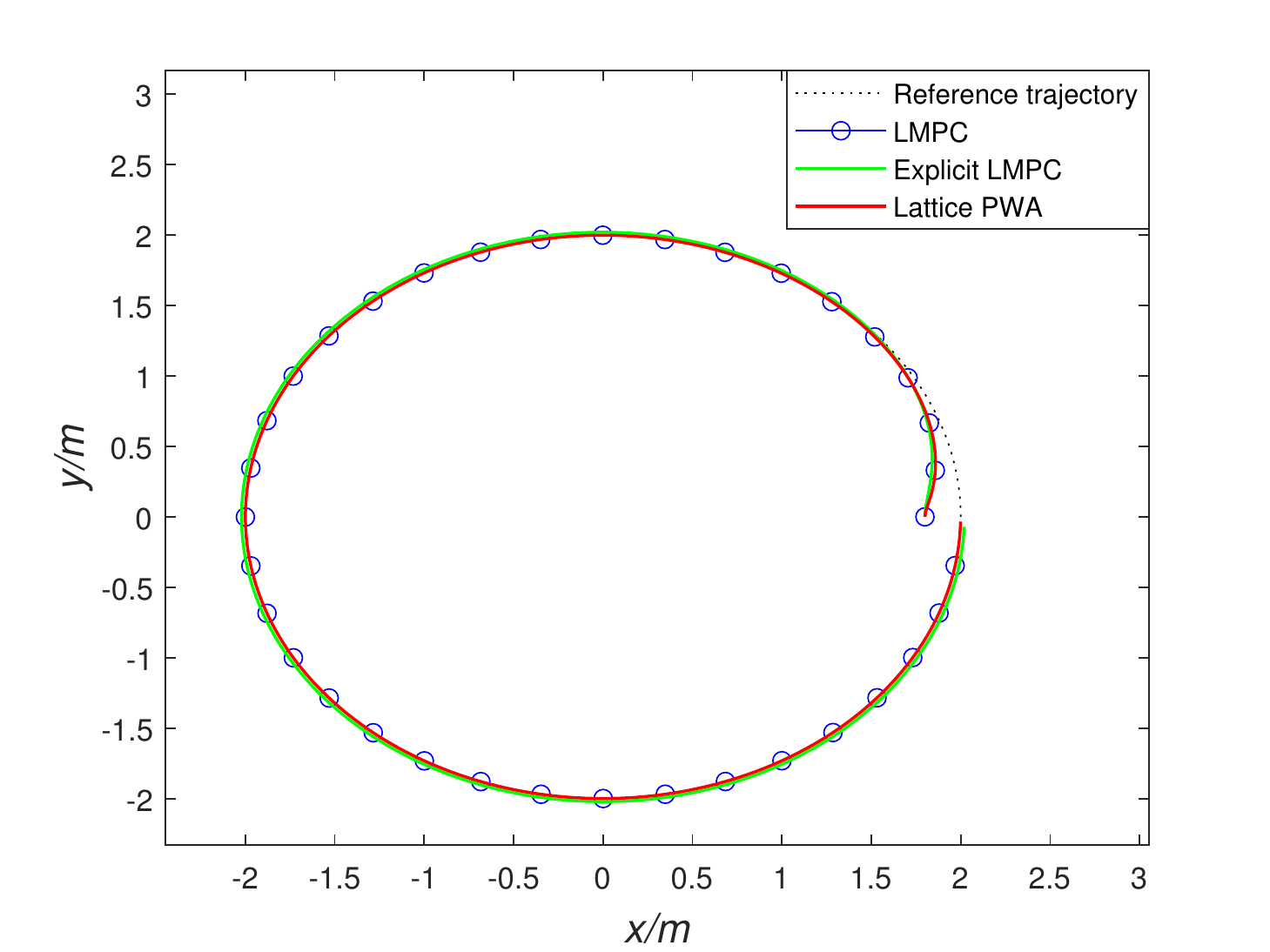} %1.png是图片文件的相对路径
			\caption{Comparison of tracking performances for a circular reference trajectory} %caption是图片的标题
			\label{Fig.5.} %此处的label相当于一个图片的专属标志，目的是方便上下文的引用
		\end{figure}

\begin{table}[htbp] 
	\centering
	\caption{Detailed tracking data for a circular trajectory} 
	\label{tab1}
	\begin{tabular}{cccc} 
		\toprule 
		Strategy & \tabincell{c}{Offline \\time (min)} & \tabincell{c}{Online \\time (ms)} & \tabincell{c}{Average tracking\\error (m)}\\
		\midrule 
		Linear MPC & $-$ & $5.5$ & $0.0043$ \\
		Explicit linear MPC & $934.68$ & $23.5$ & $0.0043$ \\
		Lattice PWA & $18.78$ & $0.056$ & $0.0043$ \\
		\bottomrule 
	\end{tabular} 
\end{table}

%Fig. \ref{Fig.5.} shows that all three methods can trace the reference trajectory, and Table \ref{tab1} gives the tracking data in detail. We can see that compared with the explicit linear MPC algorithm, the lattice PWA approximation method reduces the offline calculation time to 2.009\% and the online calculation time to 0.238\%, and the average tracking error is not significantly higher than other methods.

Fig. \ref{Fig.5.} shows that all of the methods can track the reference trajectory. The tracking data are summarized in Table \ref{tab1}. Compared with the offline and online calculation time values of the explicit linear MPC algorithm, those of the lattice PWA approximation method are reduced to 2.009\% and 0.238\%, respectively. Moreover, the average tracking error of the lattice PWA approximation method is not significantly higher than those of the other methods.

\subsection{8-shaped trajectory}
The reference trajectory is set to be in the form of the digit 8 to demonstrate the superiority of our framework. The initial state of the robot is
$\left[\begin{array}{llll}
	\!0.25\!\!& 0\!\!& 1.3\!
\end{array}\right]^{T}$.
Considering the actual situation of WMRs, the state and control quantity constraints are set as
\begin{equation}
	\begin{array}{ll}
		\left[\begin{array}{c}
			-2.5 \\
			-1.5 \\
			-2 \pi
		\end{array}\right] \leq \boldsymbol{x} \leq\left[\begin{array}{c}
			2.5 \\
			1.5 \\
			2 \pi
		\end{array}\right], \left[\begin{array}{c}
			-2 \\
			-\frac{\pi}{2}
		\end{array}\right] \leq \boldsymbol{u} \leq\left[\begin{array}{c}
			2 \\
			\frac{\pi}{2}
		\end{array}\right]
	\end{array}
\end{equation}
We obtain 252 linearized models for this trajectory and compare the tracking performances of the three methods.

\begin{figure}[h]
	\centering
	\includegraphics[width=0.5\textwidth]{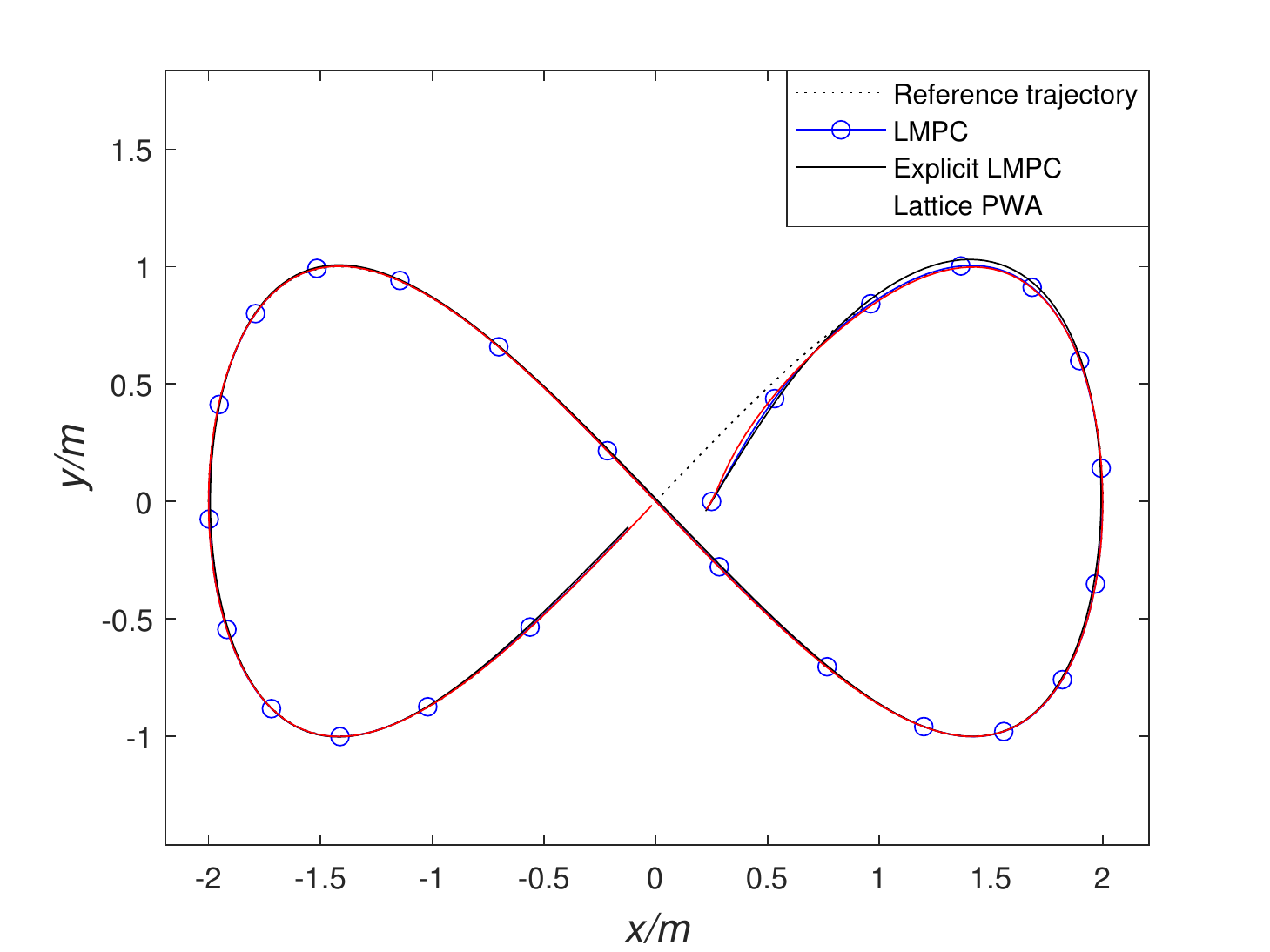} %1.png是图片文件的相对路径
	\caption{Comparison of tracking performances on an 8-shaped trajectory} %caption是图片的标题
	\label{Fig.6.} %此处的label相当于一个图片的专属标志，目的是方便上下文的引用
\end{figure}

\begin{table}[htbp] 
	\centering
	\caption{Detailed tracking data for the 8-shaped trajectory} 
	\label{tab2}
	\begin{tabular}{cccc} 
		\toprule 
		Strategies & \tabincell{c}{Offline \\time (min)} & \tabincell{c}{Online \\time (ms)} & \tabincell{c}{Average tracking\\error (m)}\\
		\midrule 
		Linear MPC & $-$ & $5.3$ & $0.0073$ \\
		Explicit linear MPC & $614.33$ & $21.5$ & $0.0073$ \\
		Lattice PWA & $11.94$ & $0.053$ & $0.0073$ \\
		\bottomrule 
	\end{tabular} 
\end{table}

%Table \ref{tab2} shows that compared with the explicit linear MPC problem, the offline calculation time of the lattice PWA approximation method is reduced to 1.944\% and the online calculation time is reduced to 0.247\% on the basis of not significantly increasing the average tracking error. The results tend to be consistent with the circular trajectory, indicating that the proposed method has better tracking performance on the whole.

Table \ref{tab2} shows that compared with the offline and online calculation time values of the explicit linear MPC problem, those of the lattice PWA approximation method are reduced to 1.944\% and 0.247\%, respectively, and the average tracking error is not significantly higher. The results are consistent with those of the circular trajectory and demonstrate that the tracking performance of our method is superior to the tracking performances of existing methods.
\section{Conclusion}
This paper proposes a lattice PWA approximation method based on explicit linear MPC to effectively track the trajectories of fast nonlinear robots. First, we successively linearize the kinematic model of the WMR along its trajectory to simplify the trajectory tracking calculations. And we get the global approximation error of the system using lattice PWA model. Second, for the explicit linear MPC problem corresponding to each linearization point, we perform lattice PWA offline to approximate the explicit control laws for decreasing the offline calculation time. In addition, the lattice PWA approximation is further simplified to reduce the online complexity and memory requirement. Finally, in the online tracking stage, the control input is calculated by substituting the actual state into the corresponding lattice PWA expression to reduce the online search time. The simulation and experimental results show that compared with the explicit linear MPC, our method exhibits superior trajectory-tracking performance, given its higher online computing speed and reduced offline computing time and memory consumption.
%\begin{thebibliography}{99}
%\nocite{*}

%\end{thebibliography}
\bibliographystyle{IEEEtran}
%\printbibliography
%\addcontentsline{toc}{section}{reference} 
%\addtolength{\bibsep}{-0.8em}
%\nocite{*}
\bibliography{reference}

\end{document}